\documentclass{ws-procs10x7}
\usepackage{balance}
\usepackage{graphicx} 

\newcolumntype{d}[1]{D{.}{.}{#1}}

\newcommand{\BsBsbar}{\ensuremath{B^0_s}-\ensuremath{\bar{B}^0_s}}

\newcommand{\Bsbar}  {\ensuremath{\bar{B}^0_s}}
\newcommand{\bs}     {\ensuremath{B_s}}

\newcommand{\keV}    {\ensuremath{\rm ke\kern -0.1em V}}
\newcommand{\MeV}    {\ensuremath{\rm Me\kern -0.1em V}}

\newcommand{\GeV}    {\ensuremath{\rm Ge\kern -0.1em V}}
\newcommand{\TeV}    {\ensuremath{\rm Te\kern -0.1em V}}

\newcommand{\ifb}    {\ensuremath{\mathrm{fb^{-1}}}}
\newcommand{\ips}    {\ensuremath{\mathrm{ps^{-1}}}}

\newcommand{\dms}    {\ensuremath{\Delta m_s}}

\newcommand{\Vtd}    {\ensuremath{V_{td}}}
\newcommand{\Vts}    {\ensuremath{V_{ts}}}

\newcommand{\VtdVts} {\ensuremath{|\Vtd/\Vts|}}
\newcommand{\Lrat}   {\ensuremath{\Lambda}}
\newcommand{\sye}[1]{\ensuremath{~\pm #1}}
\newcommand{\ase}[2]{\ensuremath{^{~+ #1}_{~- #2}}}

\newcommand{\deltaMsResultPRL}%
{\ensuremath{\dms =
  17.31\ase{0.33}{0.18}~({\rm stat})\sye{0.07}~({\rm syst})\,\ips}}
\newcommand{\deltaMsResult}%
{\ensuremath{\dms =
  17.77\sye{0.10}~({\rm stat})\sye{0.07}~({\rm syst})\,\ips}}
\newcommand{\VtdResult}%
{\ensuremath{\VtdVts =
  0.2060\sye{0.0007}~({\rm exp})\ase{0.0081}{0.0060}~({\rm theor})}}
\newcommand{\pvalue}{\ensuremath{8 \times 10^{-8}}}

\def\for{\hbox{\lowercase{for }}}

\makeindex
\begin{document}

\title{\bf \boldmath Observation of {\BsBsbar} Oscillations and\\
 measurement of {\dms} in CDF}

\author{S.~Giagu \for the CDF Collaboration}

\address{University of Rome ``La Sapienza" and Istituto Nazionale di Fisica Nucleare, Roma, Italy\\$^*$E-mail: stefano.giagu@roma1.infn.it}


\twocolumn[\maketitle\abstract{We report the observation of {\BsBsbar} 
 oscillations performed by the CDF\,II detector using a data sample of 
 1~\ifb\ of $p\bar{p}$ collisions at $\sqrt{s}=1.96~\TeV$.
 We measure the probability as a function of proper decay time that the
 {\bs} decays with the same, or opposite, flavor as the flavor at
 production, and we find a signal for {\BsBsbar} oscillations.
 The probability that random fluctuations could produce a comparable signal 
 is {\pvalue}, which exceeds $5\sigma$ significance.
 We measure {\deltaMsResult}. 
 A very important update has been presented by the CDF collaboration after 
 I gave my talk, the latest available results on {\bs} mixing are included here.}
]

\section{Introduction}
 The precise determination of the {\BsBsbar} oscillation frequency {\dms}
 from a time-dependent analysis of the {\BsBsbar} system
 has been one of the most important goals of heavy flavor 
 physics~\cite{MIXING}. This frequency can be used to strongly improve 
 the knowledge of the Cabibbo-Kobayashi-Maskawa (CKM) matrix~\cite{CKM}, and 
 to constraint contributions from new physics~\cite{UTFIT}.

 Recently, the CDF collaboration reported~\cite{CDF-BSMIX-2006} the 
 strongest evidence to date of the direct observation of {\BsBsbar} 
 oscillations, using a sample corresponding to 1~\ifb\ of data collected 
 with the CDF\,II~detector~\cite{DETECTOR_REFERENCE} at the Fermilab Tevatron.

 Here we report an update~\cite{CDF-BSMIX-2006-NEW} of this measurement 
 that uses the same data set with an improved analysis and reduces this 
 probability to {\pvalue} ($>5\sigma$), yielding the definitive 
 observation of time-dependent {\BsBsbar} oscillations. 

 The CDF analysis has been improved by increasing the {\bs} signal yield 
 and by improving the performance of the flavor tagging algorithms.
 We use {\bs} decays in hadronic
 ($\Bsbar\to D^+_s\pi^-$, $D^+_s\pi^-\pi^+\pi^-$) and semileptonic
 ($\Bsbar\to D^{+(*)}_s\ell^-\bar{\nu}_\ell$, $\ell=e$ or $\mu$)
 modes (charge conjugates are always implied), with $D^+_s$ meson 
 decaying in $D^+_s\rightarrow \phi\pi^+$, $\bar{K}^{*}(892)^{0} K^+$, and
 $\pi^+\pi^-\pi^+$, with $\phi\rightarrow K^+K^-$
 and $\bar{K}^{*0}\rightarrow K^-\pi^+$.
 We improved signal yields by using particle identification techniques to find  
 kaons from $D^+_s$~meson decays, allowing us to relax kinematic selection 
 requirements, and by employing an artificial neural
 network (ANN) to improve candidate selection.
 Signal statistics is also significantly improved by adding partially 
 reconstructed hadronic decays in which a photon or $\pi^0$ is missing:
 $\Bsbar\to D^{*+}_s\pi^-$, $ D^{*+}_s \to D^+_s \gamma/\pi^0$ and 
 $\Bsbar\to D^+_s\rho^-$, $\rho^- \to \pi^-\pi^0$,
 with $D^+_s\rightarrow \phi\pi^+$.
 Finally ANNs are used to enhance the power of the flavor 
 tagging algorithms. With all these improvements, the effective statistical 
 size of our data sample is increased respect to the previous published analysis by a factor of 2.5.

\section{Data Sample}

\begin{figure}[htb]
\begin{center}
\includegraphics[width=0.49\linewidth, height = 0.49\linewidth]{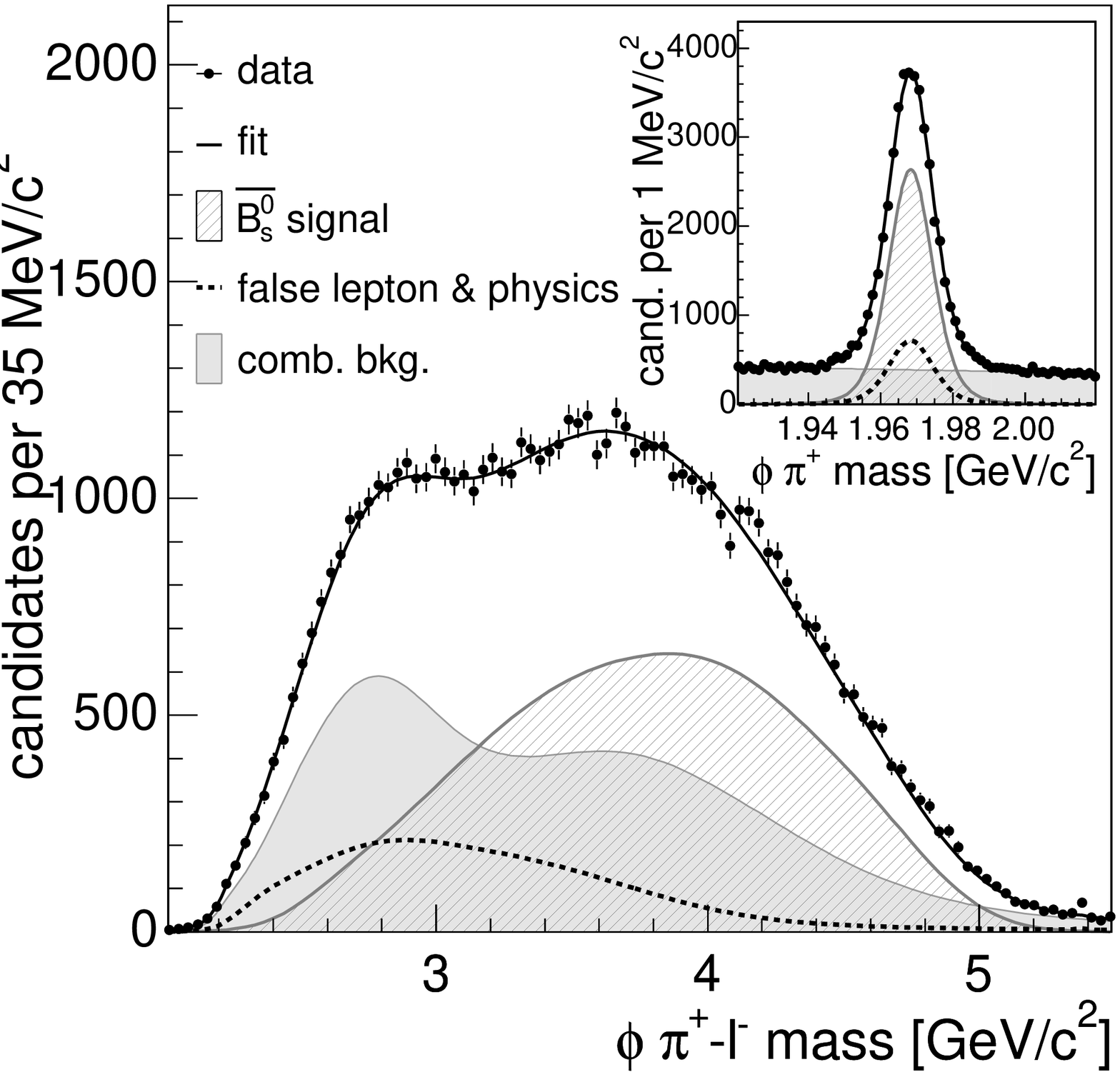}
\includegraphics[width=0.49\linewidth, height = 0.49\linewidth]{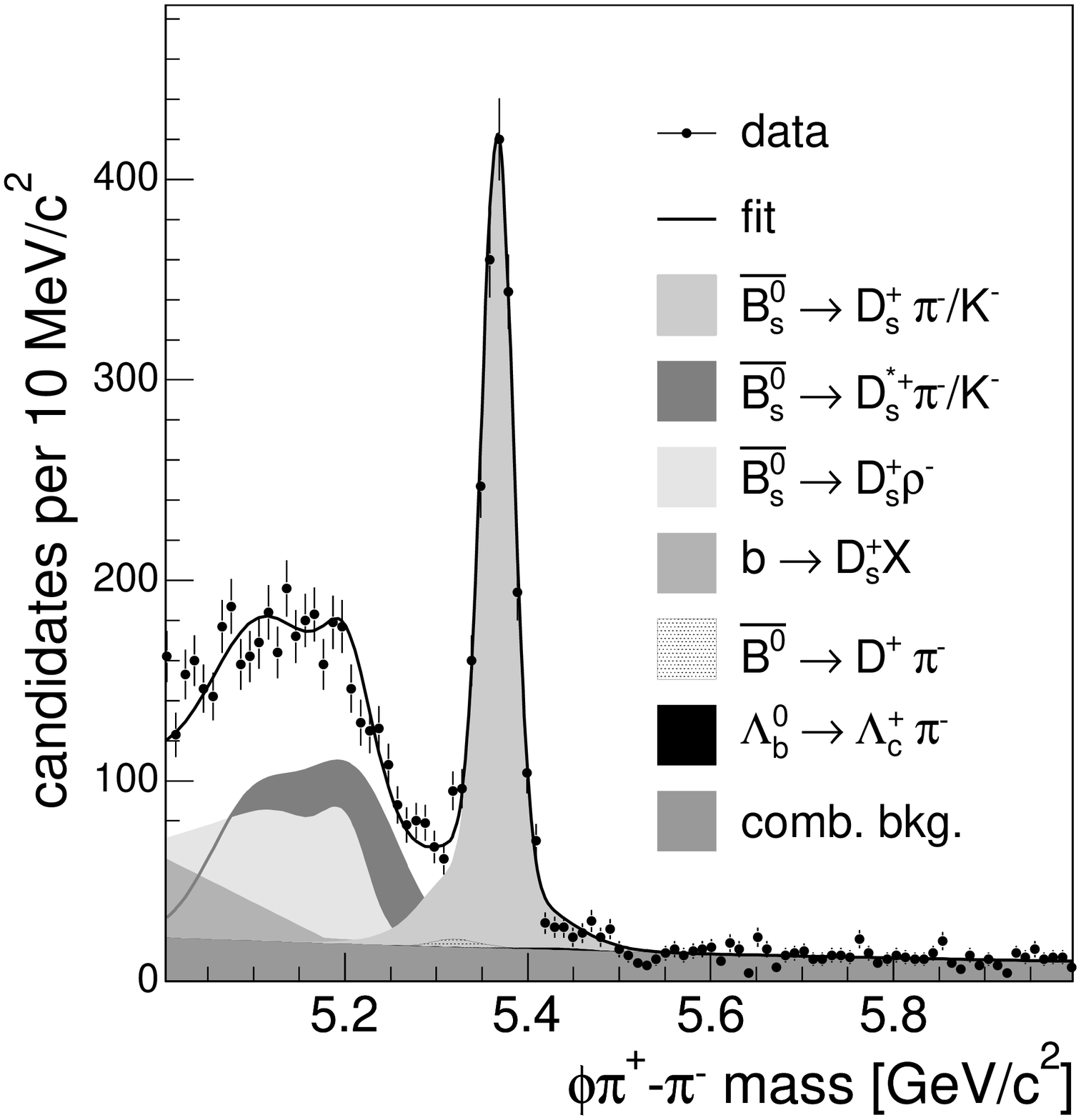}
\end{center}
\caption[] {
(Left:) The invariant mass distributions for the $D^+_s(\phi\pi^+)$
candidates [inset] and the $\ell^-D^+_s(\phi\pi^+)$ pairs.
The contribution labelled ``false lepton \& physics'' refers to backgrounds
from hadrons mimicking the lepton signature combined with real $D_s$~mesons
and other physics backgrounds.
(Right:) The invariant mass distribution for
 $\Bsbar\to D^+_s(\phi \pi^+) \pi^-$ decays including the contributions from
 partially reconstructed decays (signal contributions are drawn added on top of
 the combinatorial background).}
\label{fig:invariantmass}
\end{figure}

{\Bsbar} candidates are reconstructed by first selecting $D^+_s$ 
mesons that are lately combined with one or three additional
charged particles to form
$D^+_s\ell^-$, $D^+_s\pi^-$, or $D^+_s\pi^-\pi^+\pi^-$ candidates.
Combinatorial background is reduced by cutting on 
the minimum $p_T$ of the {\Bsbar} and its decay products,
and by requirements on the quality of the reconstructed {\Bsbar} and $D^+_s$ 
decay points and their displacement from the $p\bar{p}$ collision position.
For decay modes with kaons in the final state, a kaon identification 
variable, formed by combining TOF and $dE/dx$ information, is used to 
reduce combinatorial background from random pions or from decays from $D^+$ 
meson. The distributions of the invariant masses of the
$D_s^+(\phi\pi^+)\ell^-$ pairs $m_{D_s\ell}$ and
the $D_s^+(\phi\pi^+)$ candidates are shown in Fig.~\ref{fig:invariantmass}.
We use $m_{D_s\ell}$ to help distinguish signal, which occurs at higher
$m_{D_s\ell}$, from combinatorial and physics backgrounds.

In this analysis, we also included partially reconstructed signal
between 5.0 and 5.3\,GeV/$c^2$ from
$\Bsbar\to D^{*+}_s\pi^-$, $ D^{*+}_s \to D^+_s \gamma/\pi^0$
in which a photon or $\pi^0$ from the $D^{*+}_s$ is missing and 
$\Bsbar\to D^+_s\rho^-$, $\rho^- \to \pi^-\pi^0$ in which
a $\pi^0$ is missing.
The mass distributions for $\Bsbar\to D^+_s\pi^-$, 
$D^+_s\rightarrow \phi\pi^+$ and the
partially reconstructed signals are shown in 
Fig.~\ref{fig:invariantmass}.
Table~\ref{table:signalyields} summarizes the signal yields for 
the various decay modes. 

\begin{table}
\begin{center}
\begin{tabular}{lc}
\hline
\hline
Decay Sequence & Yield \\ \hline
$\Bsbar\to D^+_s\pi^-(\pi^-\pi^+\pi^-)$
& 5600 \\ 
$\Bsbar\to D^{+(*)}_s\ell^-\bar{\nu}_\ell$
& 61500 \\ 
Partially reconstructed
& 3100 \\ 
\hline
\hline
\end{tabular}
\caption[] {
Signal yields in hadronic, semileptonic and partially 
reconstructed decays.} 
\label{table:signalyields}
\end{center}
\end{table}

 We measure the proper decay time in the $\bs$ rest frame as
 $t = m_{\bs}L_{T}/p_T^{\rm recon}$,
 where $L_{T}$ is the measured displacement of the $\bs$ decay point with respect to
 the primary vertex projected onto the $\bs$ transverse momentum vector, and
 $p_T^{\rm recon}$ is the transverse momentum
 of the reconstructed decay products.
 In the semileptonic and partially reconstructed hadronic decays, we correct
 $t$ by a factor $\kappa = p_T^{\rm recon}/p_T(\bs)$
 determined with Monte Carlo simulation (Fig.~\ref{fig:kappa-ctres}).
 The decay time resolution $\sigma_t$ has contributions from the momentum of
 missing decay products (due to the spread of the distribution of $\kappa$)
 and from the uncertainty on $L_T$.
 The uncertainty due to the missing momentum increases with proper decay time
 and is an important contribution to $\sigma_t$ in the semileptonic decays.
 To reduce this contribution and make optimal use of the semileptonic decays,
 we determine the $\kappa$ distribution as a function of $m_{D_s\ell}$.
 The distribution of $\sigma_t$ for fully reconstructed
 decays has an average value of 87\,fs, which corresponds to one
 fourth of an oscillation period at $\dms = 17.8\,{\ips}$, and
 an rms width of $31\,{\rm fs}$.
 For the partially reconstructed hadronic decays the average $\sigma_t$
 is 97\,fs, while for semileptonic decays, $\sigma_t$ is worse due to decay
 topology and the much larger missing momentum of decay products that were not
 reconstructed (see Fig.~\ref{fig:kappa-ctres}).

\vspace{0.5 cm}
\begin{figure}[htb]
\begin{center}
\includegraphics[width=0.49\linewidth, height = 0.49 \linewidth]{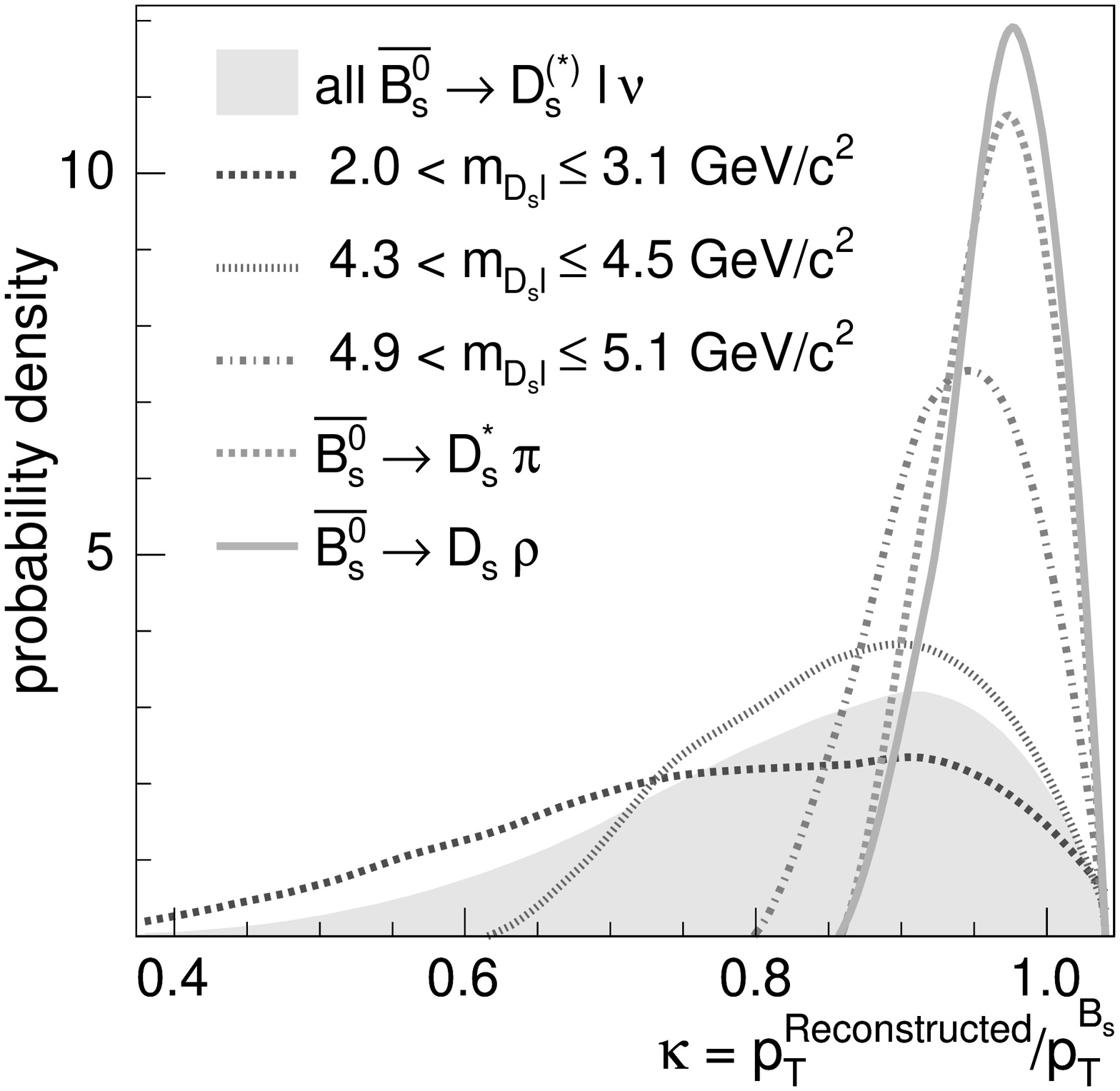}
\includegraphics[width=0.49\linewidth, height = 0.49 \linewidth]{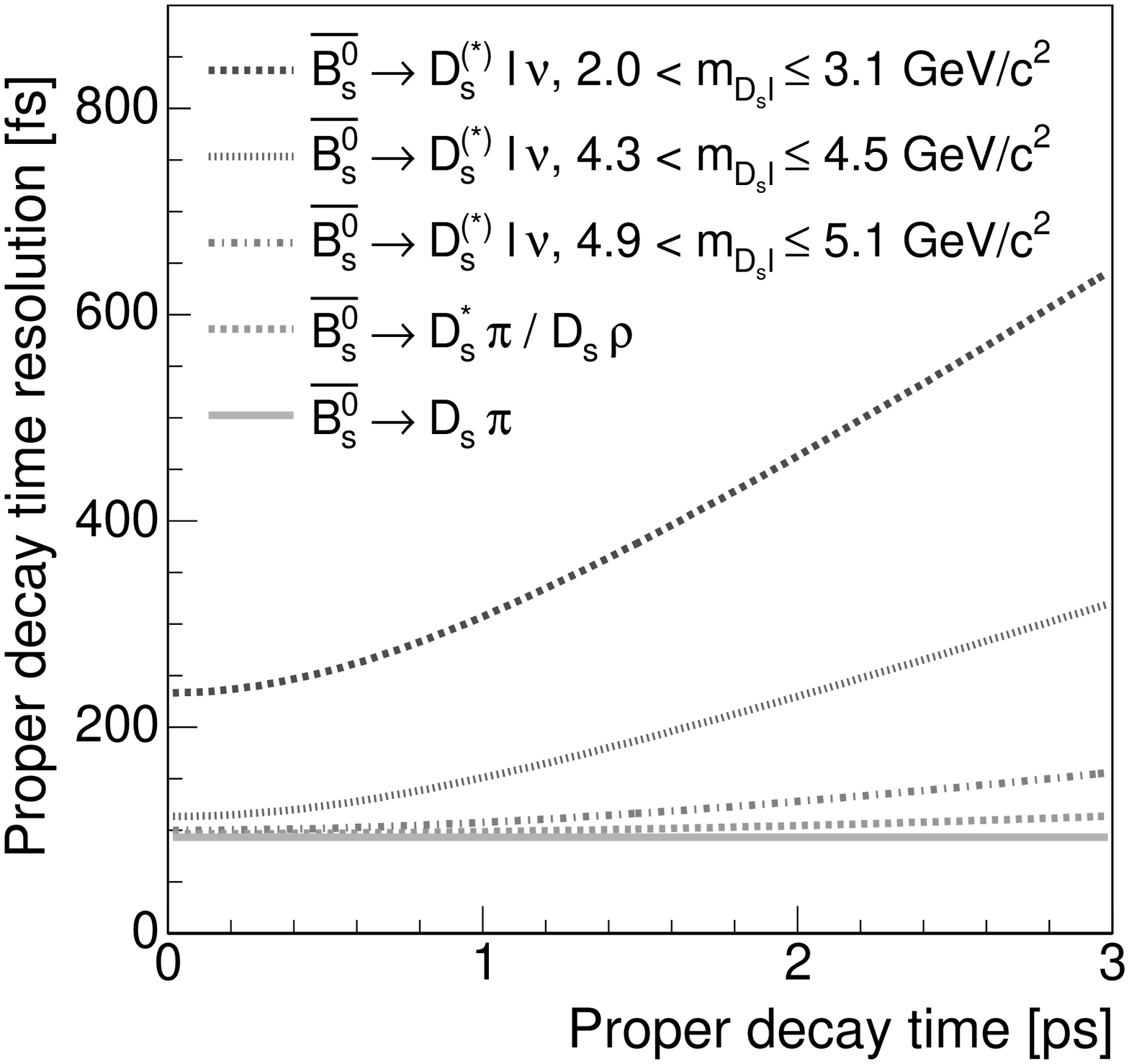}
\caption[] {
(Left:) The distribution of the correction factor $\kappa$ in
semileptonic and partially reconstructed hadronic decays from Monte
Carlo simulation.  (Right:) The average proper decay time
resolution for {\bs} decays as a function of proper
decay time.  }
\label{fig:kappa-ctres}
\end{center}
\end{figure}

\section{Flavor Tagging}

 The flavor of the {\Bsbar} at production is determined using both
 opposite-side and same-side flavor tagging techniques.
 The effectiveness $Q\equiv\epsilon {\cal D}^2$ of these techniques
 is quantified with an efficiency $\epsilon$, the fraction of signal candidates
 with a flavor tag, and a dilution ${\cal D}\equiv 1-2w$,
 where $w$ is the probability that the tag is incorrect.
 At the Tevatron, the dominant $b$-quark production mechanisms produce
 $b\bar{b}$~pairs.

 Opposite-side tags infer the production flavor of the {\Bsbar} from the
 decay products of the $b$~hadron produced from the other $b$~quark
 in the event. In this analysis we used lepton ($e$ and $\mu$) charge
 and jet charge as tags, and if both types of tag were present,
 we used the lepton tag. We also used an opposite-side flavor tag based on 
 the charge of identified kaons, and we combine the information from the 
 kaon, lepton, and jet charge tags using an ANN.
 The dilution is measured in data using large samples of $B^-$, which 
 do not change flavor, and $\bar{B}^0$,
 which can be used after accounting for their well-known oscillation frequency.
 The combined opposite-side tag effectiveness is $Q = 1.8\pm 0.1\,\%$.

 Same-side flavor tags are based on the charges
 of associated particles produced in the fragmentation of the
 $b$~quark that produces the reconstructed {\Bsbar}.
 We use an ANN to combine kaon particle-identification likelihood with
 kinematic quantities of the kaon candidate into a single tagging variable.
 Tracks close in phase space to the $\Bsbar$ candidate are
 considered as same-side kaon tag candidates, and the track with the
 largest value of the tagging variable is selected as the tagging track.
 We predict the dilution of the same-side tag using simulated data samples
 generated with the {\sc pythia} Monte Carlo~\cite{PYTHIA} program.
 Control samples of $B^-$ and $\bar{B}^0$ are used to validate
 the predictions of the simulation. The effectiveness of this flavor tag 
 is $Q = 3.7\%$ ($4.8\%$) in the hadronic (semileptonic) decay sample.
 If both a same-side tag and an opposite-side tag are present,
 we combine the information from both tags assuming they are independent.

\section{Fit and Results}

 We use an unbinned maximum likelihood fit to search for {\BsBsbar} oscillations. 
 The likelihood combines mass, decay time, decay-time resolution, and
 flavor tagging information for each candidate, and includes terms for
 signal and each type of background.
 Following the method described in \cite{MOSER}, we fit for the oscillation
 amplitude $\cal A$ while fixing $\Delta m_s$ to a probe value.
 The oscillation amplitude is expected to be consistent with
 ${\cal A} = 1$ when the probe value is the true oscillation frequency,
 and consistent with ${\cal A} = 0$ when the probe value is far from
 the true oscillation frequency.
 Figure~\ref{fig:amplitudeScan} shows the fitted value of the
 amplitude as a function of the oscillation frequency
 for the semileptonic candidates alone, the hadronic candidates alone,
 and the combination. 
 The sensitivity~\cite{CDF-BSMIX-2006,MOSER} is 19.3\,{\ips} for the
 semileptonic decays alone, 30.7\,{\ips} for the hadronic decays alone,
 and 31.3\,{\ips } for all decays combined.
 At $\dms = 17.75~\ips $, the observed amplitude
 ${\cal A} = 1.21\pm 0.20~({\rm stat.})$
 is consistent with unity, indicating that the data
 are compatible with {\BsBsbar} oscillations with that frequency, while the
 amplitude is inconsistent with zero: ${\cal A}/\sigma_{\cal A}=6.05$,
 where $\sigma_{\cal A}$ is the statistical uncertainty on ${\cal A}$
 (the ratio has negligible systematic uncertainties).
 The small uncertainty on ${\cal A}$ at $\dms = 17.75~\ips$
 is due to the superior decay-time resolution of the hadronic decay modes. 

\begin{figure}[ht]
\begin{center}
\includegraphics[width=\linewidth]{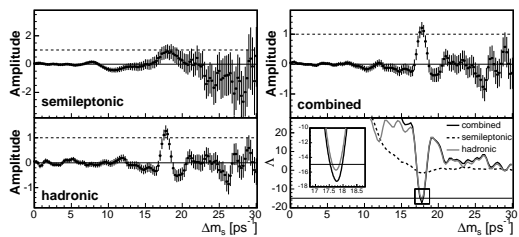}
\caption[] {
 The measured amplitude values and uncertainties versus the {\BsBsbar}
 oscillation frequency $\dms$. (Upper Left) Semileptonic decays only.
 (Lower Left) Hadronic decays only. (Upper Right) All decay modes combined. 
 (Lower Right)
 The logarithm of the ratio of likelihoods for amplitude equal to
 one and amplitude equal to zero versus the oscillation frequency.
}
\label{fig:amplitudeScan}
\end{center}
\end{figure}

 We evaluate the significance of the signal using
 $\Lrat \equiv \log [ {\cal L}^{{\cal A}=0} / {\cal L}^{{\cal A}=1}(\dms)]$,
 which is the logarithm of the ratio of likelihoods for the hypothesis
 of oscillations (${\cal A}=1$) at the probe value and the hypothesis
 that ${\cal A}=0$, which is equivalent to random production flavor tags.
 Figure~\ref{fig:amplitudeScan} shows {\Lrat } as a function of {\dms }.
 Separate curves are shown for the semileptonic data alone (dashed),
 the hadronic data alone (light solid), and the combined data (dark solid).
 At the minimum $\dms=17.77~ \ips$, $\Lrat = -17.26$.
 The significance of the signal is the
 probability that randomly tagged data would produce a value 
 of {\Lrat } lower than $-17.26$ at any value of {\dms}.
 We repeat the likelihood scan 350 million times with random tagging decisions;
 28 of these scans have $\Lambda<-17.26$, corresponding to a probability of
 $\pvalue$ ($5.4\,\sigma$), well below $5.7 \times 10^{-7}$ ($5\,\sigma$).
 
\begin{figure}[h]
\begin{center}
\includegraphics[width=0.55\linewidth]{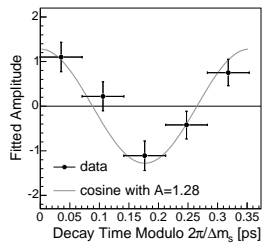}
\caption[] {
 The {\BsBsbar} oscillation signal (only hadronic decays) measured in five 
 bins of proper decay time modulo the measured oscillation period $2\pi/\dms$.
 The curve shown is a cosine with an amplitude of 1.28, which is 
 the observed value in the amplitude scan for the hadronic sample at 
 $\dms=17.77$\,$\ips$.}
\label{fig:asym}
\end{center}
\end{figure}

 To measure {\dms}, we fix ${\cal A}=1$ and fit for the oscillation frequency.
 We find {\deltaMsResult}.
 
 The only non-negligible systematic uncertainty on {\dms } is from
 the uncertainty on the absolute scale of the decay-time measurement.
 
 The {\BsBsbar} oscillations are depicted in Fig.~\ref{fig:asym} where
 candidates in the hadronic sample are collected in five bins
 of proper decay time modulo the measured oscillation period $2\pi/\dms$.

\balance

\end{document}